# Knee menisci segmentation and relaxometry of 3D ultrashort echo time (UTE) cones MR imaging using attention U-Net with transfer learning

Michal Byra[1,2], Mei Wu[1], Xiaodong Zhang[1], Hyungseok Jang[1], Ya-Jun Ma[1], Eric Y Chang[1,3], Sameer Shah[4], Jiang Du[1]

Affiliations:

[1]Department of Radiology, University of California, San Diego, CA, USA

[2]Department of Ultrasound, Institute of Fundamental Technological Research, Polish Academy of Sciences, Warsaw, Poland

[3]Radiology Service, VA San Diego Healthcare System, San Diego, USA

[4]Department of Orthopedic Surgery and Bioengineering, University of California, San Diego, CA, USA

Corresponding Author:
Jiang Du, Ph.D. e-mail: jiangdu@ucsd.edu
Department of Radiology, University of California, San Diego, USA

**Abstract**

**PURPOSE:** To develop a deep learning-based method for knee menisci segmentation in 3D ultrashort echo time (UTE) cones magnetic resonance (MR) imaging, and to automatically determine MR relaxation times, namely the T1, $T1_\rho$, and T2* parameters, which can be used to assess knee osteoarthritis (OA).

**METHODS:** Whole knee joint imaging was performed using 3D UTE cones sequences to collect data from 61 human subjects. Regions of interest (ROIs) were outlined by two experienced radiologists based on subtracted $T1_\rho$-weighted MR images. Transfer learning was applied to develop 2D attention U-Net convolutional neural networks for the menisci segmentation based on each radiologist's ROIs separately. Dice scores were calculated to assess segmentation performance. Next, the T1, $T1_\rho$, T2* relaxations, and ROI areas were determined for the manual and automatic segmentations, then compared.

**RESULTS:** The models developed using ROIs provided by two radiologists achieved high Dice scores of 0.860 and 0.833, while the radiologists' manual segmentations achieved a Dice score of 0.820. Linear correlation coefficients for the T1, $T1_\rho$, and T2* relaxations calculated using the automatic and manual segmentations ranged between 0.90 and 0.97, and there were no associated differences between the estimated average meniscal relaxation parameters. The deep learning models achieved segmentation performance equivalent to the inter-observer variability of two radiologists.

**CONCLUSION:** The proposed deep learning-based approach can be used to efficiently generate automatic segmentations and determine meniscal relaxations times. The method has the potential to help radiologists with the assessment of meniscal diseases, such as OA.

## INTRODUCTION

Osteoarthritis (OA) is the most common form of arthritis in the knee, and the menisci play an important role in the initiation and progress of OA[1]. Various menisci pathologies, such as proteoglycan loss or deterioration of the collagen network, have been directly associated with the symptomatic knee OA[2–4]. There is growing interest in developing imaging biomarkers that could help clinicians assess and monitor the progress of OA *in vivo*. Magnetic resonance (MR) imaging can provide quantitative data related to the relaxometry and morphology of the whole joint knee anatomy. For example, the meniscal T2 and T2* relaxation times were reported to be sensitive to OA-related pathophysiological processes[5,6]. Menisci, however, have a short T2 and demonstrate low signal on conventional MR sequences, making quantitative assessment with the standard clinical MR sequences infeasible. In comparison, ultrashort echo time (UTE) sequences with TEs about 100 times shorter than those of conventional sequences can be used to image tissues with short T2[7]. 3D UTE cones sequences have been proposed to measure the T1 and T1$_\rho$ relaxations for different knee tissues, including the menisci[8,9]. Given the possibility of using MR relaxations as promising OA biomarkers, assessment of the menisci still requires 3D segmentation, which is often performed manually and is therefore time-consuming and affected by inter-observer variability[10,11]. Development of robust automatic menisci segmentation methods could provide clinicians with quantitative MR parameters, such as T1, T1$_\rho$ and T2* relaxations, and would present an important step for efficient assessment and monitoring of OA.

Nowadays, deep learning methods are gaining momentum in medical image analysis[12]. These data-driven algorithms automatically process input images to learn high-level data representations and provide the desired output, such as a decision whether the investigated image contains a pathology. Deep convolutional neural networks (CNNs) have proved to be extremely useful in solving various medical image analysis problems, including image classification and segmentation[12]. The U-Net CNN and its variations are perhaps the most popular deep learning methods used for image segmentation[13]. These CNNs for segmentation consist of contracting and expanding paths with skip connections. The contracting path (encoder) processes the input image using convolutional operators to extract a compact high level image representation. In the next step, the expanding path

(decoder) utilizes this representation to generate a binary mask indicating the location in the image of the object to be segmented. The skip connections are used to include information from the contracting path in the expanding path to improve object localization. CNN-based methods were proposed for cartilage segmentation[14–16] and whole joint anatomy segmentation, including menisci[17]. 2D U-Net CNNs for menisci segmentation were developed based on double-echo steady-state knee joint MRI images collected from healthy subjects and from patients with OA[18,19]. The authors demonstrated that automatic segmentations could be used to assess menisci morphology and to extract efficient biomarkers for OA diagnosis[18,19]. Additionally, automatic menisci segmentation methods based, for example, on fuzzy logic and extreme learning machines were developed before deep learning techniques gained their momentum[20–22]. Moreover, semi-automatic menisci segmentation methods based on region growing were developed; those methods proved to perform well in the case of OA biomarker extraction[23–25]. Based on the obtained segmentations, the authors calculated meniscal $T1_\rho$ and T2 parameters, and related those to the levels of OA progression[24]. However, the quantitative evaluation of menisci (e.g., T1, $T1_\rho$, T2, and T2* relaxation times) is challenging due to the lack of signal with conventional gradient echo sequences with echo times around 4 to 7 ms, which are too long to accurately quantify meniscus with a short T2* of around 5 ms[26].

Due to small medical datasets, it is a common practice to use transfer learning and utilize a pre-trained deep learning model. This way, a model developed using a large dataset can be adjusted to address the medical imaging problem of interest. The better performing CNNs for classification, such as the VGG19 or InceptionV3[27,28], have been developed using the ImageNet dataset, which includes over 1,000,000 RGB images[29]. The first convolutional layers of these deep CNNs identify low level concepts in the images, such as color blobs, illustrating the importance of these basic features for efficient object recognition[30]. In practice, it is usually reasonable to utilize these already developed convolutional operators for the new deep learning model. In MR imaging, transfer learning has been proven to provide good results in, for instance, image classification[31]. In the study on whole joint anatomy segmentation, the authors pre-trained their CNN using a different set of MR images[17]. In the case of computer vision, it was demonstrated that transfer learning methods could improve the CNN-based road scene semantic segmentation[32].

The aim of this work is to show the feasibility of automatic quantitative characterization of the menisci in 3D UTE cones MR imaging. We present an "end to end" method for the automated segmentation of the menisci and the extraction of quantitative MR parameters, namely the T1, T1$_\rho$, and T2* relaxation times. First, we use transfer learning to develop an attention 2D U-Net CNN based on a relatively small set of MR volumes acquired using 3D UTE cones sequences. For this task, we utilize a model pre-trained on a large set of non-medical images. Additionally, we employ the attention mechanism to improve the segmentation performance. Second, we compare the average T1, T1$_\rho$, and T2* values calculated using manual and automatic segmentations. The usefulness of our approach is evaluated using regions of interest (ROIs) provided by two experienced radiologists. We study the inter-observer variability between the radiologists and evaluate the CNNs developed using ROIs provided by each radiologist. To our knowledge, this study presents the first examples of knee menisci segmentation using transfer learning with deep convolutional neural networks in 3D Cones MR imaging for accurate assessment of meniscal relaxometry.

## METHODS

### UTE imaging and data collection

A total of 61 human subjects (aged 20-88 years, mean age 55±16 years; 30 males, 31 females) was recruited for this retrospective study. Initial clinical screening included a clinical exam and lower extremity radiograph to select patients with no knee OA symptoms as well as patients with suspicion of OA. Informed consent was obtained from all subjects in accordance with guidelines of the University of California San Diego Institutional Review Board. The study included 27 healthy participants, 25 patients with mild OA, and 13 patients with moderate OA. Whole knee joint imaging was performed using 3D UTE-Cones sequences on a 3T MR750 scanner (GE Healthcare Technologies, Milwaukee, WI). An 8-channel knee coil was used for signal excitation and reception. The protocol included 3D UTE Cones imaging and measurement of T1, T1$_\rho$, and T2* relaxations. The basic 3D UTE-Cones sequence employed a short rectangular pulse for signal excitation, followed by k-space data acquisition along twisted spiral trajectories ordered in the form of multiple cones. T1 was quantified using 3D UTE-Cones with actual flip angle imaging (AFI) and variable flip angle (VFA) approach, where B1 inhomogeneity was mapped using the 3D

UTE-Cones-AFI technique, followed by accurate T1 mapping using the 3D UTE-Cones-VFA technique. T1$_\rho$ was quantified using 3D UTE-Cones-AdiabT1$_\rho$ imaging, where identical non-selective AFP pulses with a duration of 6.048 ms, bandwidth of 1.643 kHz, and maximum B1 amplitude of 17 µT were used to generate T1$_\rho$ contrast[8]. Multispoke acquisition after AdiabT1$_\rho$ preparation was incorporated for improved time-efficiency (e.g., N$_{sp}$ spokes were acquired per adiabatic T1T1$_\rho$ preparation). T2* was quantified by acquiring fat-saturated multi-echo UTE-Cones data. All 3D UTE Cones data were acquired with a field of view of 15×15×10.8 cm$^3$ and receiver bandwidth of 166 kHz. Other sequence parameters were: 1) 3D UTE-Cones-T$_2$*: TR=45 ms; FA=10°; matrix=256×256×36; fat saturation; multi-echo of 0.032, 4.4, 8.8, 13.2, 17.6, and 22 ms; and scan time of 3 min 40 sec; 2) 3D UTE-Cones-AFI: TR$_1$/TR$_2$=20/100 ms; flip angle=45°; matrix=128×128×18; and scan time of 4 min 57 sec; 3) 3D UTE-Cones-VFA: TR=20 ms; flip angle=5°, 10°, 20°, and 30°; matrix=256×256×36; and scan time of 9 min 28 sec; 4) 3D UTE-Cones-AdiabT$_{1\rho}$: TR=500 ms; FA=10°; matrix=256×256×36; N$_{sp}$=25; number of AFP pulses N$_{AFP}$=0, 2, 4, 6, 8, 12, and 16; each with scan time of 2 min 34 sec. The total acquisition time for the four UTE-Cones sequences was approximately 35 minutes. The Levenberg-Marquardt algorithm was used for non-linear fitting of UTE-Cones data based on prior reported equations to calculate T1, T1$_\rho$, and T2*[33]. All calculations were performed in Matlab (Mathworks, Natick, MA).

To account for potential motion during the relatively long acquisitions, the elastix motion registration based on the Insight Segmentation and Registration Toolkit was applied to the 3D UTE-Cones data before quantification[34,35]. The first set of UTE data (UTE-Cones-T2* data) was treated as fixed images, and the remaining sets of data (AFI, VFA, and AdiabT$_{1\rho}$) were treated as moving images. 3D non-rigid registration was applied to register the moving images to fixed images. In the 3D non-rigid registration, both rigid (affine) and non-rigid (B-spline) were applied as a two-staged approach to register the images. All registrations were driven by Advanced Mattes mutual information[34]. The transformations were obtained by registration of the grayscale images (source UTE images), which were then applied to the labeled images. Adaptive stochastic gradient descent optimizer was used to optimize both the affine and B-spline registration.

Two experienced radiologists with 22 (Rad 1) and 14 (Rad 2) years of experience participated in our study. First, we investigated which images would be best for outlining the ROIs. Menisci show as high signal in the UTE images, but as lower signal with later echoes. Subtraction of a later echo image from the first echo may provide high contrast imaging of menisci. Different subtracted images were reconstructed and presented to the radiologists in order to select those providing the best visibility of the menisci in respect to the surrounding tissues. Based on subjective assessment, subtracted AdiabT1ρ-weighted MR images corresponding to $N_{AFP}$ of 0 and 2 were used to outline the ROIs independently by both radiologists (Fig. 1). Additionally, the lateral meniscus (LM) and medial meniscus (MM) were indicated.

**Model development and performance evaluation**

The deep learning approach employed in our study was based on the U-Net architecture, see Fig. 2. Similar models, all inspired by this architecture, achieved good results in the case of the menisci segmentation in the previous papers[17–19]. Additionally, we employed attention layers to process the feature maps propagated through the skip connections[36,37]. Self-attention mechanisms proved to improve segmentation performance in the case of small objects in computed tomography[37]. Clearly, the menisci constitute a small part of the knee and, consequently, the whole MR image. Attention layers help the network focus more on small regions, instead of analyzing the entire field of view. In the case of the standard U-net architecture, feature maps from the encoder path are directly concatenated with the output of the decoder convolution layers. This output is related to menisci localization in the image. In our case, the attention layers filter the encoder feature maps based on the output of the decoder convolution layers (Fig. 2) to incorporate the information about initial menisci localization. Therefore, areas of feature maps that are far from the initial menisci localization are compressed. This way, less noisy feature maps are propagated through the skip connections. Moreover, we employed the following transfer learning-based approach to the model development: The weights of the first two convolutional blocks of our U-Net were initiated with the weights of the corresponding first two convolutional blocks of the VGG19 model pre-trained on the ImageNet dataset[27,29]. The first layers of pre-trained CNNs like the VGG19 commonly include blob and edge detectors; therefore, these layers can provide generic image features useful for the analysis of, for instance, MR images, which are similar to natural images. Moreover, deep layers extract high-level features more related to the particular recognition problem. Transfer learning methods utilizing the VGG19 CNN performed well for various medical image analysis problems across different medical imaging modalities[12,31,38]. Thanks to this replacement, our U-Net CNN demonstrated its capability to extract low level image features from the beginning. These capabilities did not have to be redeveloped using the training set. Another issue was related to the fact that the VGG19 CNN was trained using RGB images as input. MR images, on the other hand, are grayscale, which raises a question about how to efficiently utilize the pre-trained model. The most common approach is to duplicate the grayscale intensities across all color channels[31,39]. Moreover, the grayscale images should be normalized accordingly in order to use the pre-trained network more efficiently. This issue is important since we use the subtracted images, which have a unique pixel intensity distribution. Generally, the optimal normalization could be determined using the validation set. However, this would be time consuming. To address the mentioned issues automatically, we decided to add an additional 1D convolutional block (with a bias term) consisting of three convolutional filters to the front of our U-Net. This way, the images are first processed before they are passed to the convolutional blocks originating from the VGG19 CNN. The aim of this block is to adjust the subtracted images, rescale the pixel intensities, and perform grayscale-to-RGB conversion in order to more efficiently utilize the power of the pre-trained convolutional blocks. The parameters of this layer can be determined during the training with the backpropagation algorithm[40].

In this work, in comparison to the previous studies[17–19], we decided to train our network using the Dice score-based loss function. The Dice score (or coefficient) is defined in the following way:

$$\text{Dice score} = \frac{2\,|M \cap A|}{|M| + |A|},$$

where *M* is the manual segmentation ROI, *A* is the automatic segmentation ROI predicted by the CNN, and $|\cdot|$ refers to set cardinality. Usage of the Dice score for training has several advantages. First, this metric is commonly used to assess the segmentation performance; therefore, the maximization of this parameter is desirable. Second, many studies show that the Dice score-based training is a good choice for heavily imbalanced data[41,42], such as objects, like menisci, that occupy a small part of the whole image.

The dataset was divided into training, validation, and test sets with a 36/10/15 split. The validation set included data from 5 healthy participants and from 5 patients with mild or moderate OA. The test set included data from 7 healthy participants and from 8 patients with mild or moderate OA. Next, the 3D MR volumes were broken down into 2D images. In our case, working with 2D data had several advantages. First, it enabled the usage of transfer learning with deep models pre-trained on natural 2D images. Second, we avoided training a deep model for 3D MRI images, which would have required large amounts of data. Third, models trained on 2D MR images are usually more general; for example, changing the number or distance between MR image slices does not influence the model. Only the 2D MR images containing the meniscus were used for the development and evaluation of the models. Contrast of each image was improved using Matlab implementation of the edge-aware contrast manipulation algorithm with the parameters selected experimentally[43]. Next, the images and the corresponding binary masks were automatically cropped to 192x192 to narrow the field of view, then resized to the default VGG19 input size of 224x224. The MR images were resized using the bilinear transformation, whereas for the binary masks, the nearest neighborhood algorithm was applied. The training set consisting of 458 2D images was augmented by image rotation and horizontal flipping to produce 2748 images. During the training, we monitored the Dice score on the validation set. The U-Net was trained using the backpropagation with the Adam optimizer[44]. Weights of the layers were initialized using the Xavier uniform initializer[45]. The batch size was set to 32. The learning rate and the momentum were set to 0.001 and 0.9, respectively. However, the learning rate was exponentially decreased every five epochs by using a drop factor of 0.5 if no improvement was observed on the validation set. The training was stopped if no improvement in respect to the Dice Score was observed on the validation set after 15 epochs. After the training, the better performing model with respect to the validation set was selected. The networks were trained in Python using Tensorflow[46]. The experiments were performed on a computer equipped with four GeForce

GTX 1080 Ti graphics cards. By CNN 1 and CNN 2, we refer to the models trained separately using the ROIs provided by the first and the second radiologist, respectively.

After the training, the better performing CNN models were employed to calculate the ROIs using the test set, which contained 191 2D images from 15 menisci. In the next step, the manual and automatic segmentations were used to calculate the average T1, $T1_\rho$, and T2* relaxations for each 2D ROI. Average percentage relative absolute distance errors between the raters and the CNNs were calculated. Due to small number of patients with different levels of OA, we calculated the errors using the entire dataset. Two-sided $t$-test at the significance level of 0.01 with the Bonferroni correction was applied to examine whether the mean estimates were significantly different in the case of the manual and automatic segmentations. Dice scores, Pearson's linear correlation coefficient, and Bland-Altman plot were used to assess the level of agreement between the CNNs and radiologists. We separately compared the results obtained for the entire menisci, LM, and MM. We examined four cases, Rad 1 vs Rad 2, Rad 1 vs CNN 1, Rad 2 vs CNN 2, and CNN 1 vs CNN2, respectively. Moreover, for each CNN, we determined the receiver-operating characteristic (ROC) curve, then calculated the areas under the ROC curve (AUC) to assess how effective the networks were at detecting menisci pixels[47]. To assess the robustness of the proposed method, we applied the CNNs to the MR images from the test set with no menisci present. This was done to assess whether the CNNs might generate false positives. In this case, an image was classified to have menisci if the CNN detected at least two adjacent menisci pixels.

**RESULTS**

The average Dice score for the ROIs outlined by two radiologists was equal to 0.820, indicating good inter-observer agreement, see Table 1. Both CNNs produced good Dice coefficients of 0.860 and 0.833 for the Rad 1 vs CNN 1 and the Rad 2 vs CNN 2, respectively. Both CNNs were excellent at detecting menisci pixels, with the AUC values equal to 0.96 and 0.95 for the CNN 1 and CNN 2, respectively. The CNNs did not generate false positives when applied to the test slices with no menisci present. The Dice score between the Rad 1 and CNN 2 (developed using the second radiologist's ROIs) was equal to 0.835. Similarly, in the case of the Rad 2 and CNN 1, the Dice score was equal to 0.818.

These results show that the agreement between the radiologists is similar to agreement between one radiologist and the deep learning model developed using ROIs provided by another radiologist. Moreover, the Dice score for the ROIs produced by the CNNs was equal to 0.882, being significantly higher than the score obtained for two radiologists ($p$-value<0.001). In the case of the MM and LM assessed separately, we obtained similar results. However, as is presented in Table 1, the Dice scores obtained for the MM segmentations were slightly higher in each case.

Table 2 lists the average values of the T1, $T1_\rho$, T2*, and ROI areas calculated based on the manual and automatic segmentations. There were no associated differences between the relaxation parameters estimated using the manual and automatic segmentations; for the T1, the $p$-values were equal to 0.96 and 0.65 for the CNN 1 and CNN 2, respectively. The corresponding $p$-values for the $T1_\rho$ estimation were equal to 0.71 and 0.49. Similar results were obtained for the T2* estimation, with $p$-values of 0.55 and 0.91, respectively. In the case of the ROI area estimation, the second radiologist outlined significantly larger ROIs on average than the first ($p$-value<0.001), as depicted in Table 2. However, the ROIs calculated using the CNN 1 did not differ significantly from the ROIs produced by the CNN 2 ($p$-value=0.07). Moreover, there was no difference between the manual and automatic segmentations with respect to the ROI area calculations, with $p$-values of 0.50 and 0.68, respectively. The average percentage relative absolute distance errors for the CNN 1 and CNN 2 were equal to 1.95% and 2.26% for T1, 2.56% and 3.03% for $T1_\rho$, 4.59% and 6.15% for T2*, and 14.27% and 15.21% for the ROI area estimation, respectively.

Strong Dice scores indicated good agreement between the quantitative parameters estimated using the manual and automatic segmentations, as depicted in Figs. 3, 4, and 5. In Fig. 3, the Pearson's linear correlation coefficients for the T1 values for the Rad 1 vs Rad 2, Rad 1 vs CNN 1, Rad 2 vs CNN 2, and CNN 1 vs CNN 2 were equal to 0.91, 0.94, 0.90, and 0.95, respectively. For the $T1_\rho$ values depicted in Fig. 4, the coefficients were equal to 0.92, 0.96, 0.93, and 0.95, respectively. Finally, for the T2* relaxations shown in Fig. 5, the linear correlation coefficients were equal to 0.92, 0.97, 0.94, and 0.97, respectively. All obtained correlation coefficients were high ($p$-values<0.001), indicating a good level of agreement between the radiologist and the deep learning models, see Bland-Altman plots in Figs. 3, 4, and 5. Fig. 6 shows the manual and automatic segmentations obtained for four cases from the test set. Figs. 6a and b illustrate cases in which a high level of agreement was achieved, while Figs. 6c and d present examples in which a lower level of agreement between the radiologists and radiologists/models was obtained.

**DISCUSSION**

In this work, we proposed an efficient deep learning-based approach to knee menisci segmentation and extraction of quantitative MR parameters in 3D UTE Cones MR imaging. ROIs provided by two radiologists were used as the ground truth for the performance evaluation. Our results show that the proposed approach can successfully segment the menisci and can provide quantitative MR measures (relaxation times and morphology) with similar accuracy to those obtained by two radiologists. The level of agreement, measured by the Dice score, between the radiologists was similar as between the radiologists and the deep learning models developed using ROIs provided by another radiologist. The Dice score between the ROIs generated by the two radiologists was equal to 0.820, while the Dice score between the first radiologist and CNN 2 (developed using the second radiologist's ROIs) was equal to 0.839. In the case of the second radiologist and CNN 1, the Dice score was equal to 0.818. Additionally, both CNNs produced strong Dice coefficients equal to 0.860 and 0.835 for the CNN 1 and CNN 2, respectively. These results illustrate an important issue related to the assessment of the segmentation algorithms: the Dice score of 1.0 in the case of the CNN 1 would result in the Dice score of 0.820 for the Rad 2 vs CNN 1 comparison because, in this case, the ROIs generated by the network would be the same as those provided by the first radiologist. Ideally, the inter-observer agreement should be taken into account to evaluate the performance of the automatic segmentation in a fairer way. The Dice score between the ROIs generated using the CNN 1 and CNN 2, equal to 0.882, was significantly higher than the Dice score for the radiologists, which was 0.820. This promising result suggests that the performance of the automatic segmentation was driven more by the underlying image data and that the deep learning may provide results less affected by the inter-observer variability of the ROI annotators.

The Dice scores achieved by the proposed method are comparable or better than the results reported in the previous papers[17–19]. In [18], the authors used a 2D U-Net developed from scratch on a set of two different datasets. The first set included 464 T1$_\rho$-weighted volumes and, in this case, a Dice score of approximately 0.685 was obtained (we averaged the Dice scores reported in the paper for the LM and MM to simplify the comparison). The second set included 174 3D double-echo steady state volumes; for this dataset, the Dice score of approximately 0.771 was achieved. In [19], the authors employed a 2D U-Net trained using 88 3D double-echo steady state volumes from the Osteoarthritis Initiative and achieved a Dice score of around 0.828. In this case, however, the ROIs calculated by the 2D U-Net were additionally processed using a 3D U-Net to further improve the results. Using this method, the Dice score of 0.888 was obtained. In [17], the authors used a U-Net-inspired CNN to segment whole joint knee anatomy, including menisci. The model was trained with 20 3D data volumes acquired using sagittal frequency fat-suppressed 3D fast spin-echo sequence. In this case, the CNN model was first pre-trained using a 60 sagittal 3D T1-weighted spoiled gradient recalled echo knee image dataset which included ROIs of cartilage and bone outlined by imaging experts[48]. The authors achieved a Dice score of around 0.72 for the menisci segmentation, which, similar to [19], was subsequently improved using additional post-processing algorithms, producing the final Dice score of 0.831. Although the Dice scores obtained in our study were high, direct comparison of our results with the Dice scores obtained by other authors is difficult for several reasons. First, in the mentioned studies, different MR data were used to develop the networks. As presented in [18], there was a significant difference in performance between the models developed using T1$_\rho$-weighted images and the models developed using 3D double-echo steady state volumes. For the menisci segmentation, we employed MR images obtained through UTE MR imaging. Images generated in this way, however, have lower quality in comparison to regular clinical images. It remains to be investigated which MR imaging method should be used to obtain the best segmentation results. The approach proposed in our study may serve as a first step in answering this question. In our case, several different images of the menisci were presented to radiologists and they were asked to select the one that would be best for preparing the ROIs. Comparison of the results is also difficult due to the fact that the methods described in previous papers were developed based on reference ROIs provided by radiologists and technicians with different levels of experience. In [49], results from a large number of biomedical image analysis challenges, including segmentation, were analyzed, demonstrating that making a change to the reference annotations may change the challenge winner's ranking; therefore, segmentation performance is related to the quality of annotations. In our study, the CNNs trained using ROIs provided by two radiologists achieved different Dice scores. Moreover, except for the results reported in [18], the authors developed the algorithms without a real validation set, which could result in overfitting and overly optimistic Dice scores. In the case of our study, the dataset was divided into training, validation, and test sets, and the model selection was performed using the validation set.

Our study shows the feasibility of providing quantitative MR measures for the menisci in an automatic fashion. There were no statistical differences between the average T1, $T1_\rho$, and T2* parameters calculated for each ROI using the manual and automatic segmentations. Additionally, we obtained good linear correlation coefficients of at least 0.90 for the manual and automatic segmentations. It is worth noting that, as in the case of the Dice score, the agreement between the CNNs was higher than that of the radiologists. The correlation coefficient for the T2* estimation was equal to 0.97, while the correlation coefficient for the radiologists was equal to 0.92. As far as we know, this is the first study reporting T1, $T1_\rho$, and T2 relaxation values of the menisci based on fully automatic segmentation of 3D UTE images. Nevertheless, we can compare our results with those reported in one of the previous studies on cartridge compartments segmentation[18]. The authors compared $T1_\rho$ and T2 values determined using ROIs generated by a radiologist and 2D U-Net, and obtained good linear correlation coefficients of around 0.92.

The proposed approach to model development has several advantages. First, thanks to the transfer learning and attention mechanism, we were able to develop a well-performing 2D U-Net CNN for menisci segmentation. The first two convolutional blocks of our U-Net were replaced with the blocks extracted from the VGG19 CNN. In comparison to the transfer learning method employed in one of the previous studies[17], where the CNN was pre-trained on a set of 60 MR knee data volumes, our approach utilized convolutional blocks of a deep network trained using over 1,000,000 images. The first convolutional blocks of the deep CNNs commonly contain blob and edge detectors, which are crucial for successful object recognition. Due to the transfer learning, our U-Net did not have to redevelop operations responsible for edge extraction during the training. Additionally, we introduced a small convolutional block in front of the first pre-trained block to better match the grayscale MR images, since the VGG19 CNN was originally developed for RGB images. Nevertheless, there are several issues related to our approach. First, we developed the CNNs using a relatively small dataset of MR images. Generally, the performance of deep learning algorithms is expected to increase with the data volume; thus, it would be better to have a large annotated dataset to train the model from scratch. However, annotating (or, in our case, outlining the ROIs) large datasets is usually time consuming and sometimes impractical; therefore, for applications, it is rather desirable to develop as efficient a model as possible using the smallest, yet most optimal dataset. A second issue is related to the applied transfer learning technique, which was developed for 2D images. Because of this, it is not straightforward to apply the proposed technique in this paper's transfer learning method to 3D cases. Another issue is that the radiologists were asked to select the best possible images for annotations. This selection was done in a strictly subjective way. The subtracted images selected for the menisci segmentation might not be optimal for other knee joints segmentation, but this requires further study.

In the future, following the study presented in [17], we would like to develop a deep learning model for the segmentation of the whole knee anatomy. Based on this segmentation, it would be possible to calculate the quantitative MR parameters for each knee joint tissue. In several papers, the segmentation provided by the 2D U-Net was consequently processed using, for example, a 3D U-Net[14,19] to further improve the results. While in our study we obtained an accuracy equivalent to the inter-observer variability of two radiologists, it would be interesting to explore the post-processing possibilities in order to further improve the results and make the segmentation more robust. Aside from developing the segmentation networks, we would also like to continue collecting data from human subjects, especially from those with mild and moderate OA, and assess the usefulness of automatically extracted MR quantitative parameters for classification of patients with different levels of OA.

**CONCLUSIONS**

In this paper, we presented an efficient deep learning-based approach to menisci segmentation and extraction of quantitative parameters in 3D ultra-short echo time cones magnetic resonance imaging. The method, evaluated using regions of interest provided by two radiologists, demonstrated efficacy with respect to segmentation performance and magnetic resonance parameters determination. The proposed transfer learning-based segmentation algorithm achieved performance similar to that obtained by the radiologists. The results and techniques presented in our study may serve as an important step to providing magnetic resonance biomarkers for osteoarthritis diagnosis and monitoring.

## ACKNOWLEDGMENT

The authors acknowledge grant support from GE Healthcare, NIH (1R01 AR062581, 1R01 AR068987 and 1R01 NS092650), and the VA Clinical Science Research & Development Service (1I01CX001388, I21RX002367).

## CONFLICT OF INTEREST STATEMENT

The authors do not have any conflicts of interest.

**Table 1**. The average Dice scores (plus median and 95% CI) calculated using the manual and automatic segmentations. Rad 1 and Rad 2 refers to ROIs outlined by two radiologists. CNN 1 and CNN 2 indicates the ROIs generated by the CNNs developed using the ROIs provided by the first and second radiologist, respectively. MM – medial meniscus, LM – lateral meniscus.

|  |  | Rad 1 | Rad 2 | CNN 1 | CNN 2 |
|---|---|---|---|---|---|
| Rad 1 | MM+LM | 1 | - | - | - |
|  | MM |  |  |  |  |
|  | LM |  |  |  |  |
| Rad 2 | MM+LM | 0.820 (0.839, 0.807-0.831) | 1 | - | - |
|  | MM | 0.826 (0.842, 0.809-0.843) |  |  |  |
|  | LM | 0.814 (0.832, 0.797-0.830) |  |  |  |
| CNN 1 | MM+LM | 0.860 (0.871, 0.850-0.868) | 0.818 (0.845, 0.805-0.831) | 1 | - |
|  | MM | 0.872 (0.882, 0.862-0.883) | 0.831 (0.857, 0.812-0.850) |  |  |
|  | LM | 0.847 (0.884, 0.833-0.862) | 0.805 (0.832, 0.787- 0.823) |  |  |
| CNN 2 | MM+LM | 0.835 (0.850, 0.825-0.844) | 0.833 (0.863, 0.819-0.846) | 0.882 (0.900, 0.873-0.890) | 1 |
|  | MM | 0.841 (0.855, 0.829-0.853) | 0.841 (0.864, 0.822-0.861) | 0.894 (0.903, 0.885 0.904) |  |
|  | LM | 0.829 (0.847, 0.815-0.843) | 0.825 (0.855, 0.806-0.843) | 0.871 (0.896, 0.857-0.885) |  |

**Table 2**. The average quantitative MR parameters (plus median and 95% CI) calculated using the manual and automatic segmentations. Rad 1 and Rad 2 refers to ROIs outlined by two radiologists. CNN 1 and CNN 2 indicates the ROIs generated by the CNNs developed using the ROIs provided by the first and second radiologist, respectively. MM – medial meniscus, LM – lateral meniscus.

|  |  | Rad 1 | Rad 2 | CNN 1 | CNN 2 |
|---|---|---|---|---|---|
| T1 [ms] | MM+LM | 959.4 (961.8, 947.9-970.9) | 967.0 (979.9, 954.8-979.3) | 958.9 (960.3, 947.5-970.4) | 963.1 (962.6, 951.6-974.6) |
|  | MM | 969.9 (961.6, 955.3-984.6) | 973.5 (975.9, 957.1-989.9) | 964.4 (957.3, 950.3-978.5) | 967.6 (957.9, 953.2-981.9) |
|  | LM | 947.5 (961.8, 929.5-965.4) | 959.7 (980.9, 941.2-978.3) | 952.7 (966.3, 934.2-971.3) | 958.1 (975.9, 939.5-976.7) |
| $T1_\rho$ [ms] | MM+LM | 27.5 (27.4, 26.9-28.0) | 28.0 (28.0, 27.5-28.6) | 27.3 (27.4, 26.8-27.9) | 27.8 (27.8, 27.2-28.3) |
|  | MM | 27.9 (28.0, 27.1-28.8) | 28.5 (28.5, 27.7-29.2) | 27.7 (27.8, 27.0-28.5) | 28.0 (28.0, 27.3-28.7) |
|  | LM | 26.9 (26.6, 26.1-27.7) | 27.5 (27.3, 26.7-28.4) | 26.9 (26.6-26.1-27.6) | 27.5 (27.4, 26.7-28.3) |
| T2* [ms] | MM+LM | 9.8 (9.0, 9.3-10.3) | 9.9 (9.0, 9.4-10.4) | 9.6 (9.0, 9.1-10.0) | 9.9 (9.1, 9.4-10.4) |
|  | MM | 9.6 (9.0, 9.0-10.1) | 9.7 (9.1, 9.2-10.3) | 9.4 (9.1, 8.9-9.9) | 9.6 (9.2, 9.1-10.1) |
|  | LM | 10.1 (8.9, 9.2-10.9) | 10.1 (8.8, 9.3-10.9) | 9.7 (8.7, 8.9-10.5) | 10.2 (9.0, 9.3-11.0) |
| Area [mm$^2$] | MM+LM | 60.6 (57.5, 57.4-63.9) | 71.8 (66.1, 67.5-76.1) | 64.0 (59.5, 60.6-67.3) | 68.8 (63.5, 64.7-72.8) |
|  | MM | 65.7 (63.3, 60.6-70.9) | 79.2 (76.6, 72.4-85.9) | 68.6 (65.2, 63.1-74.0) | 73.4 (68.6, 66.8-79.8) |
|  | LM | 55.0 (52.5, 51.4-58.7) | 63.7 (61.2, 59.2-68.3) | 58.9 (56.2, 55.4-62.3) | 63.8 (60.0, 59.3-68.3) |

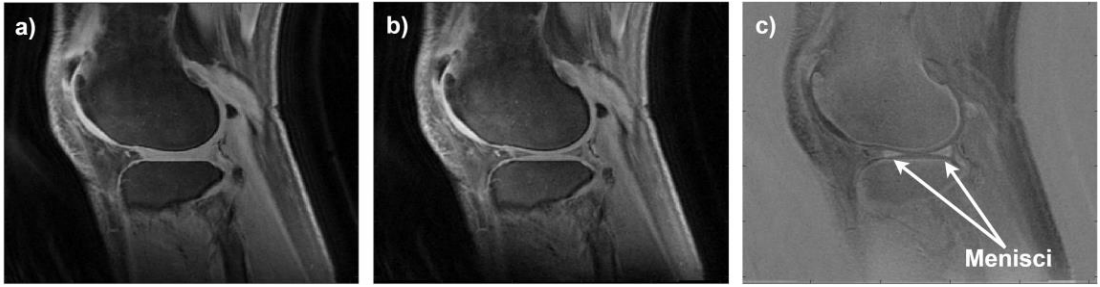

Figure 1. The MR images obtained using the UTE 3D cones for the NIR value of a) 0 and b) 2, and c) the resulting subtracted image. The subtracted images were selected by the radiologists to outline the menisci.

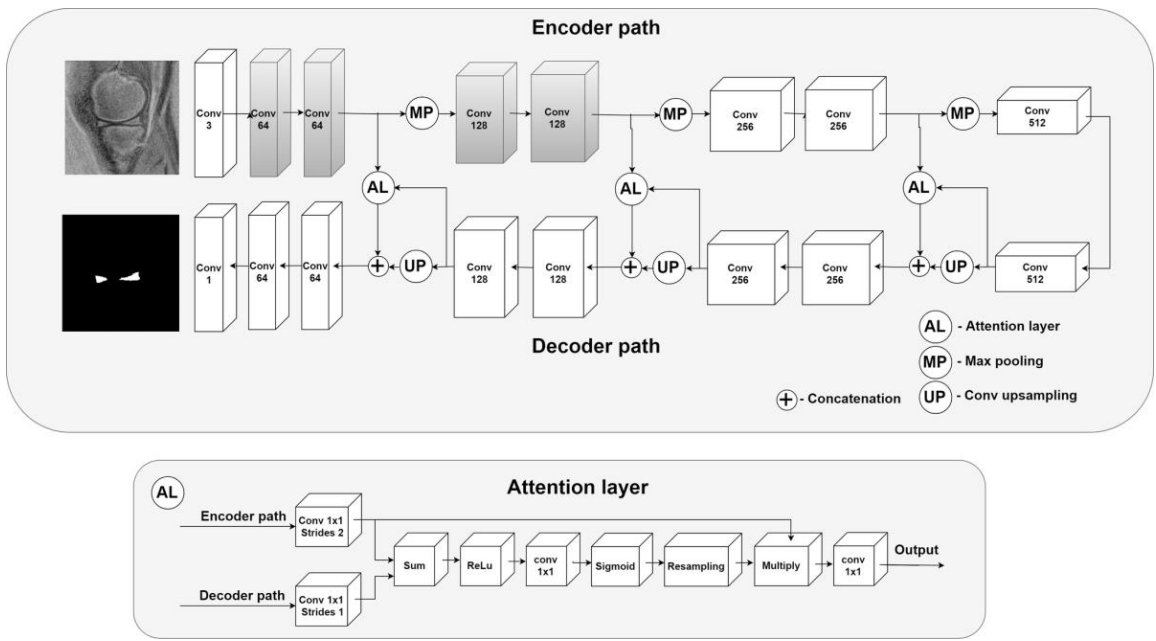

Figure 2. The proposed 2D attention U-Net CNN for the menisci segmentation. Gray colors indicate the convolutional blocks initiated with the weights extracted from the VGG19 network. For each block the number of filters is indicated below the block type. AL – attention layer, Conv – 2D convolutional block, Max pool – max pooling operator, Up – up sampling with a 2D transposed convolutional block (kernel size of 2x2, stride of 2x2). Each convolutional block, except for the first and the last block, used the rectifier linear unit (ReLu) as the activation function and 3x3 convolutional filters. The first utilized 1D 1x1 convolutional filters and no activation function was employed for this layer. The last block utilized the sigmoid activation function suitable for the binary classification. AL layers were applied to process the feature maps propagated through the skip connections, to let the network focus more on particular regions in feature maps, instead of analyzing the entire image representations.

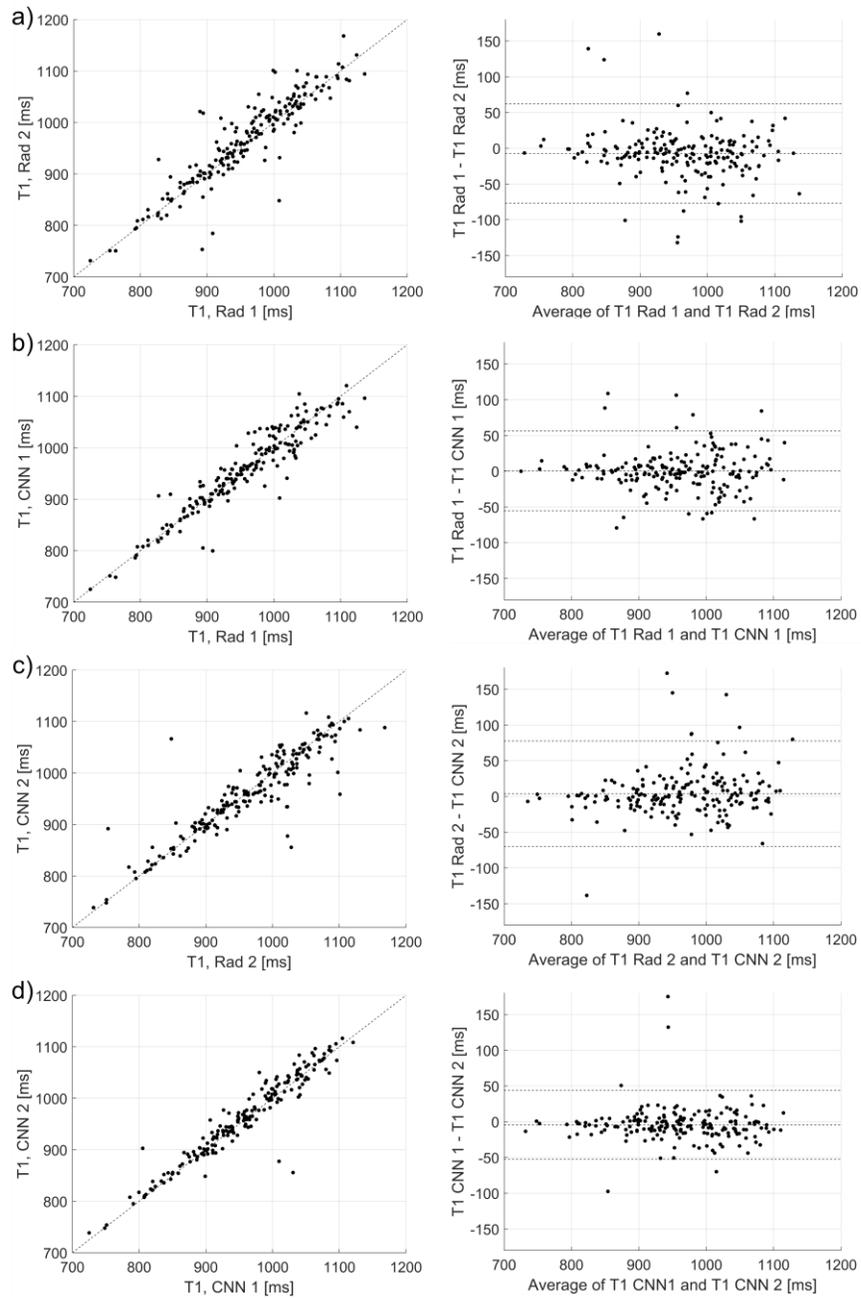

Figure 3. The relationships and Bland-Altman plots for the average T1 values calculated using the manual and automatic segmentations provided by the radiologists and the CNNs, a) relationship for the radiologists' ROIs (linear correlation coefficient of 0.91, p-value<0.001), b) relationship between the CNN1 and CNN2 (0.95, p-value<0.001). c) and d) show the corresponding relationships for the CNN 1 (0.94, p-value<0.001) and CNN 2 (0.90, p-value<0.001) in respect to the radiologists' ROIs, which were used to develop the model.

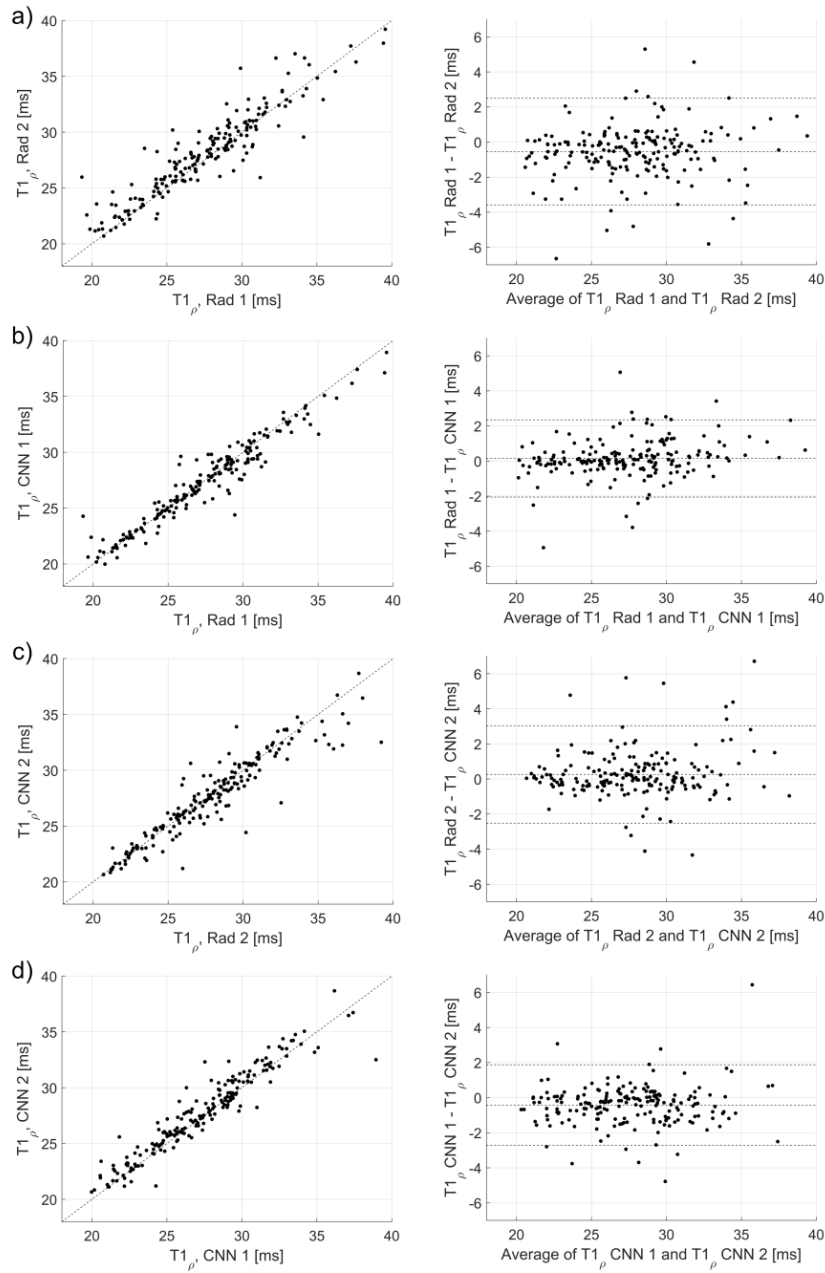

Figure 4. The relationships and Bland-Altman plots for the average T1ρ values calculated using the manual and automatic segmentations provided by the radiologists and the CNNs, a) relationship for the radiologists' ROIs (linear correlation coefficient of 0.92, p-value<0.001), b) relationship between the CNN1 and CNN2 (0.95, p-value<0.001). c) and d) show the corresponding relationships for the CNN 1 (0.96, p-value<0.001) and CNN 2 (0.93, p-value<0.001) in respect to the radiologists' ROIs, which were used to develop the model.

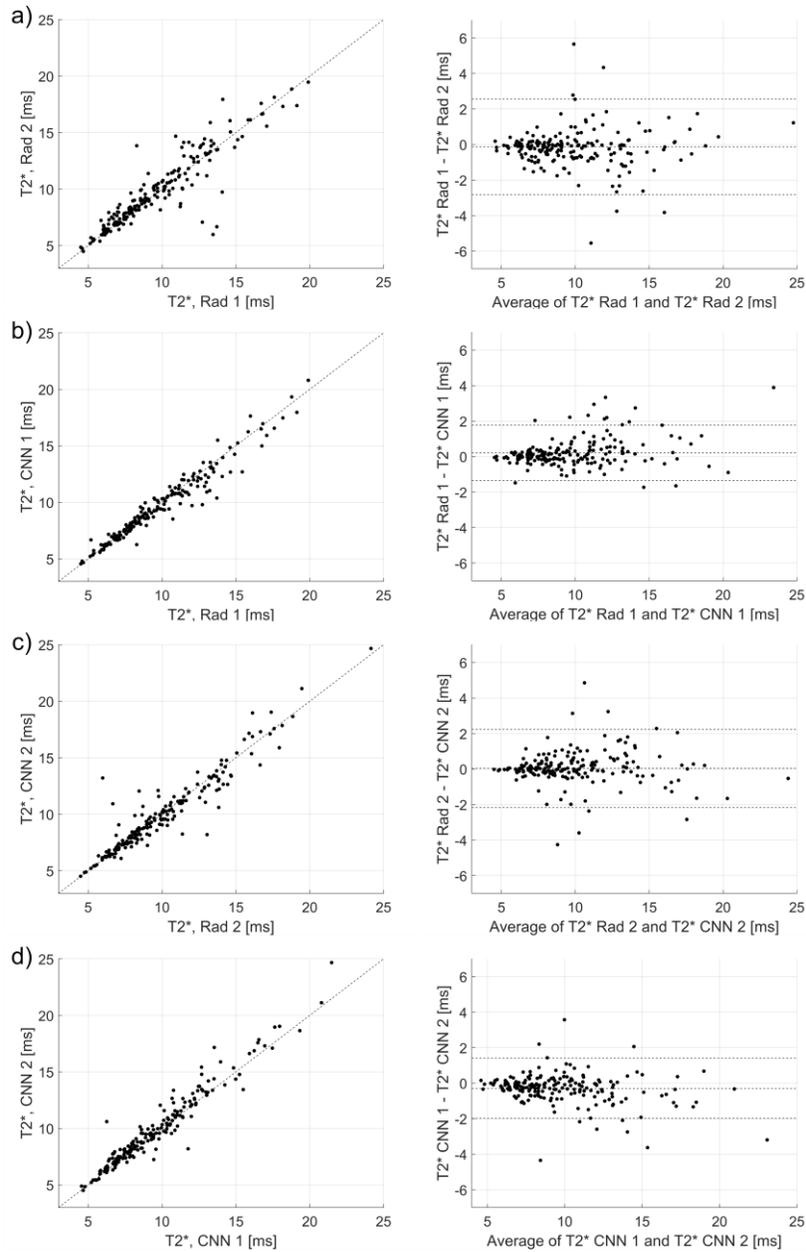

Fig. 5. The relationships and Bland-Altman plots for the average T2* values calculated using the manual and automatic segmentations provided by the radiologists and the CNNs, a) relationship for the radiologists' ROIs (linear correlation coefficient of 0.92, p-value<0.001), b) relationship between the CNN1 and CNN2 (0.95, p-value<0.001). c) and d) show the corresponding relationships for the CNN 1 (0.97, p-value<0.001) and CNN 2 (0.95, p-value<0.001) in respect to the radiologists' ROIs, which were used to develop the model.

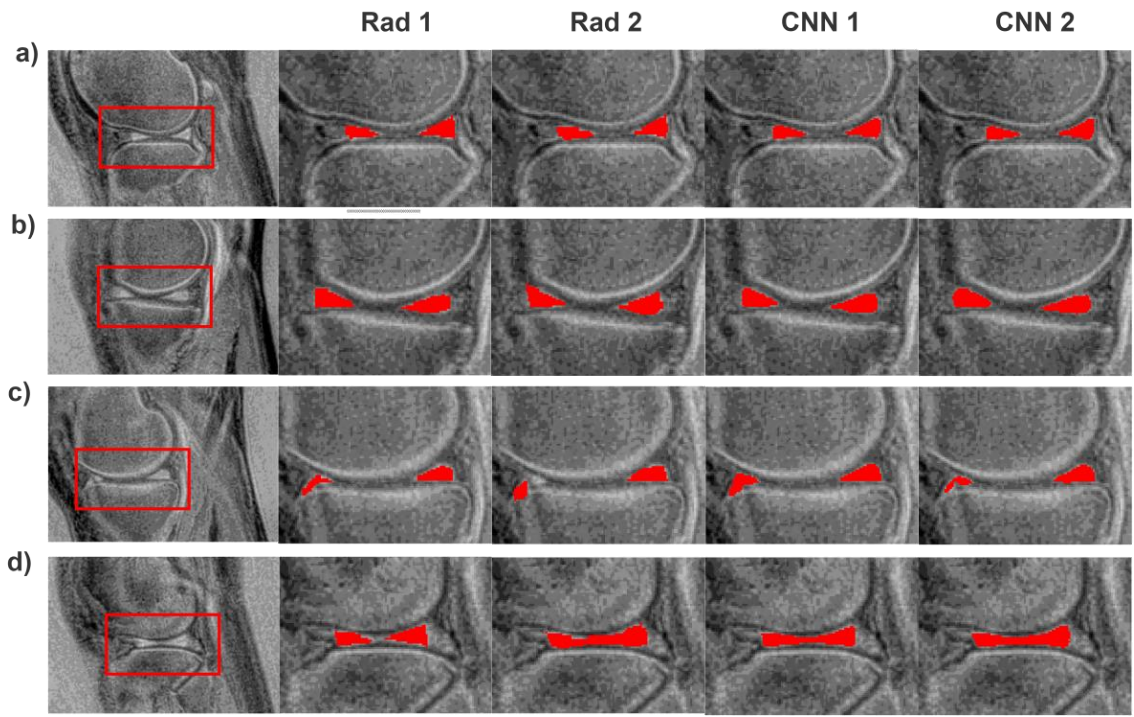

Figure 6. Examples illustrating the results of the manual and automatic segmentations. a) and b) show results for which a large level of agreement was obtained for each method. In c) the second radiologist, in comparison to the first one, outlined the right part of meniscus differently, but both models pointed out this part of the image as corresponding to the meniscus. For the images presented in d) the first radiologist outlined the meniscus in a conservative manner. In comparison, the second radiologists and both CNNs generated larger ROIs.